\pdfoutput=1
\documentclass[10pt]{article}
\usepackage{geometry}
\usepackage[hidelinks]{hyperref}
\usepackage{graphicx}
\usepackage{comment}
\usepackage{xspace}
\usepackage{color}
\usepackage[hang,flushmargin]{footmisc}
\usepackage{todonotes}
\usepackage[normalem]{ulem}
\usepackage{epstopdf}
\usepackage{wrapfig}
\usepackage{subcaption}
\usepackage{csquotes}
\usepackage{amsmath}
\DeclareMathOperator*{\argmax}{arg\,max}

\RequirePackage{amsmath,amsfonts,amssymb}
\RequirePackage{graphicx,xcolor}
\RequirePackage{booktabs}
\newcommand{\eg}{{\em e.g.},\xspace}

\begin{document}

\title{Artificial Intelligence for %Precision 
Digital Agriculture at Scale: Techniques, Policy, and Challenges}
\author{Somali Chaterji$^\xi$\footnote{$\xi$: Corresponding Author}, Nathan DeLay, John Evans, Nathan Mosier, Bernard Engel,
Dennis Buckmaster, \\
\& Ranveer Chandra}

% make the title area
\maketitle

%\noindent \underline{More at}:
%\textbf{http://bit.ly/digitized-Ag}
%\bigskip

\noindent \textbf{Abstract.}
\noindent The biggest promise of digital agriculture is the ability to evaluate the system on a holistic basis at multiple levels (individual, local, regional, and global) and generate tools that allow for improved decision making in every sub-process. 
% SC (11/10/19): I don't think the above statement makes sense. 
Recent advances in the Internet of Things (IoT) hardware and software makes it possible to collect data from diverse sources in a so called ``smart farm''. By interconnecting these IoT devices, it is possible to collect data from a large connected area at different time scales, including in near real-time (i.e., delays of a few tens of seconds). IoT devices can connect with each other in a wireless sensor network (WSN) setting and are capable of sensing different kinds of information. In row crop systems, data is generated from a variety of sources. Field operations are a large generator of data. In most systems, the majority of the data is generated in six operations, namely, \textit{soil sampling, fertilizer application, planting, scouting, spraying, and harvesting}.

\textbf{Strategic importance of digital agricultural solutions.}
Digital agriculture---encompassing precision agriculture, data analytics, and data ownership---has the promise to transform agricultural throughput. 
% SC (11/10/19): Does the digital agriculture definition encompass only these three? They seem rather disparate and at different levels of abstraction. 
It can do this by applying data science for mapping input factors to crop throughput and that too in a region-specific and crop-specific manner, while bounding the available resources. In addition, as the data volumes and varieties increase with the increase in sensor deployment in agricultural fields, data engineering techniques will also be instrumental in collection of distributed data as well as distributed processing of the data. These have to be done such that the latency requirements of the end users and applications are satisfied.
%is an important focus area for the Colleges of Agriculture and Engineering and the Department of Agricultural and Biological Engineering (ABE) and the Department of Agricultural Economics.
%SC(06/20/2019): I removed the Purdue-specific points from the line and made the abstract more generic, motivated by the global needs.
Leveraging the need for increased crop productivity for feeding the masses, Purdue's Colleges of Engineering and Agriculture are looking to transform the Midwestern counties surrounding Purdue University and specifically consisting of ten counties in the area into a hub for digital agricultural innovation. At the same time, Microsoft has developed and is looking to spread the reaches of the FarmBeats program~\cite{farmbeats-2017}, which has the vision of empowering farmers with low-cost digital agriculture solutions using low-cost sensors, drones, and computer vision and machine learning (ML) algorithms. 
% This is perfectly in sync with the land grant mission of Purdue University, highlighted under the ML and sustainability ambit of Purdue's 150 giant leaps vision. 
Understanding how farm technology and big data can improve farm productivity can significantly increase the world's food production by 2050 in the face of constrained arable land and with the water levels receding. While much has been written about digital agriculture's potential, little is known about the economic costs and benefits of these emergent systems. 
% SC (11/10/19): What emergent systems? 
In particular, the on-farm decision making processes, both in terms of adoption and optimal implementation, have not been adequately addressed. There are important questions to be answered before we can leap into data analytics for agriculture---questions related to technical viability, economic feasibility, and data protection and ownership. These questions cannot be looked upon in isolation. For example, if some algorithm needs data from multiple data owners to be pooled together, that raises the question of data ownership. This paper is the first one to bring together the important questions that will guide the end-to-end pipeline for the evolution of a new generation of digital agricultural solutions, driving the next revolution in agriculture and sustainability under one umbrella. 
% SC (11/10/19): This claim is too broad. I think this article is on data systems in digital Ag?
%The team comprising of experts in digital agriculture, applied data analytics, and agricultural policy is uniquely positioned to write this article, which will hopefully energize the community to search for answers in the right direction. 

\textbf{Identified gaps motivating this vision paper.}  
One of the key bottlenecks in leveraging the promise of digital agriculture is the lack of \textit{proven} benefit from data sharing by farmers (or ranchers in the case of livestock farming.  There are multiple factors that result in so-called adoption challenges, namely: data ownership concerns, economics, financial incentives of data ownership and dissemination, anticipated and quantified return on investment, data aggregation and pipelining from the source(s) to the desired locations, to name a few.

If even a subset of the gaps described above can be circumvented using a mix of technology, policies, and awareness, the possible outcomes of digital agriculture can shine. Some of these include seed-variety mapping to performance characteristics resulting in better selection and even engineering of seed varieties using precision technologies such as genome editing; better understanding of regional and temporal conditions leading to sustainable and localized modeling of soil supplementation needs; and capability for logistical modeling of the system, leading to increased efficiency and machine selection. Thus, we look to present both the bottlenecks and challenges on the one hand and some possible technological and policy solutions on the other to transform some of the promises of digital agriculture into a reality.

\section{Introduction}
By 2050, the world's population will increase to 9 billion, which will aggravate the food-water-energy nexus challenges. Demand will also rise because of increase in people's wealth resulting in higher meat consumption plus the increasing use of cropland for biofuels. Site-specific farm management (precision farming) has the potential to nourish the world while increasing farm profitability under constrained resource conditions. Despite advancements in field sensors, the global positioning system (GPS), and grid soil sampling, adoption of technology by farm operators has fallen short of expectations~\cite{schimmelpfennig2016farm, bullock2009value, baerenklau2007dynamics}. Moreover, it is unclear how profitable the adoption of such technologies will be~\cite{bullock2000agronomic}.
Operator demographics, operation size, and perceived benefits influence the decision to invest in site-specific management practices and adoption rates vary widely across technology types~\cite{thompson2019farmer, schimmelpfennig2016farm}. Use of variable rate technology (VRT), for example, has lagged that of yield monitors and automated guidance systems. An important distinction has been drawn between farm technology and the farm information that complements the technology~\cite{bullock2009value}. 

\begin{wrapfigure}{R}{0.65\textwidth}
\begin{center}
\includegraphics[width=0.62\textwidth]{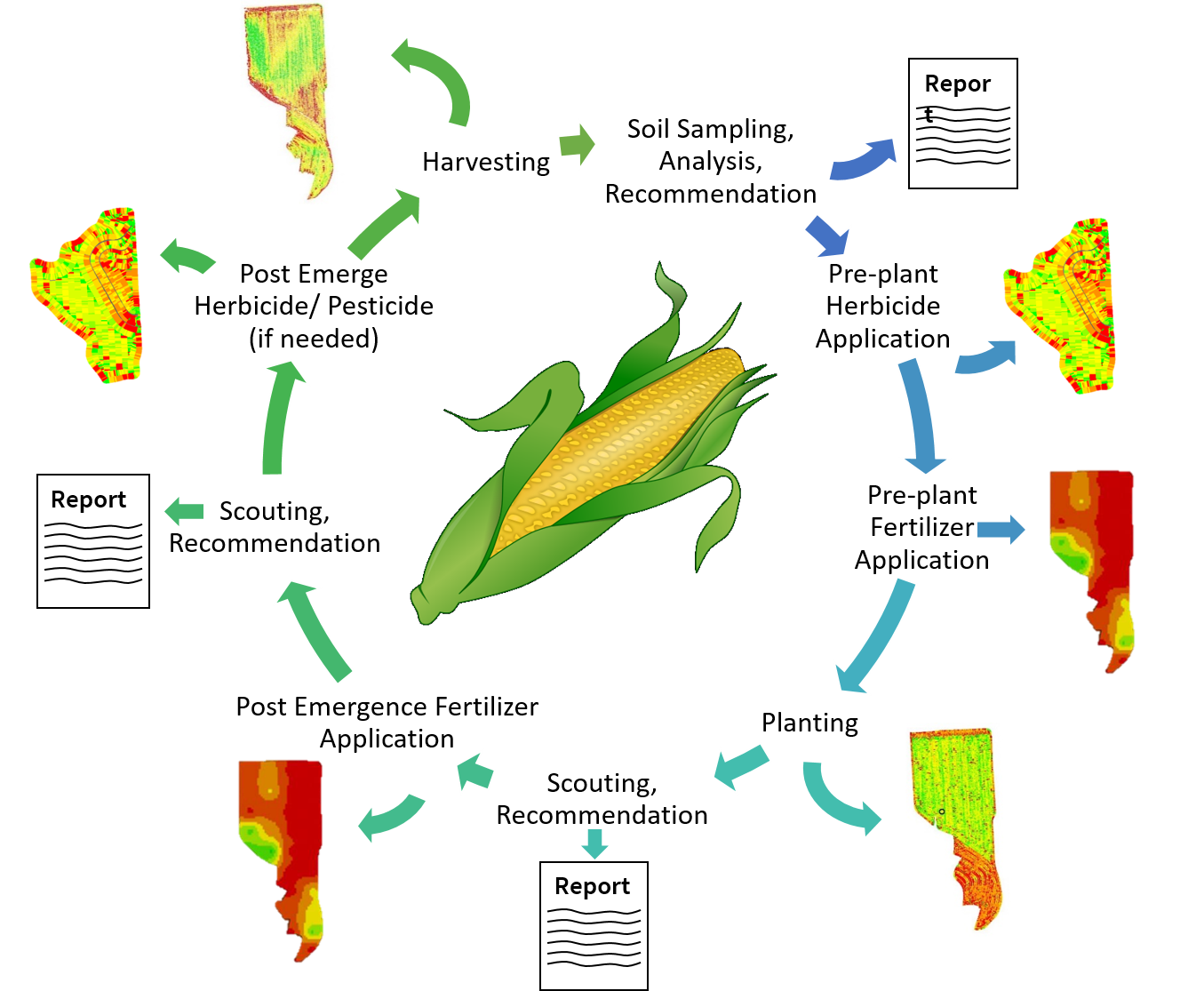}
\vspace{-0.30cm}
\caption{Example of no-till corn operation, data generation, and lifecycle. In modern agriculture, practices generate data during almost every operation. This data may be site-specific or more broadly defined, but often there is no clear way to aggregate data layers.}
\label{fig:ag-data}
\end{center}
\vspace{-0.30in}
\end{wrapfigure}

For example, the economic benefit of a variable rate application technology (VRT) for fertilizer treatment depends on accurate intra-field soil data, which is itself expensive to obtain. Unless the economic returns to site-specific management cover both the up-front investment and the cost of collecting quality information, adoption will be low. Specifically, for the deployment of VRT for fertilizer application, the following are some of the guidelines: marking management zones for being entered into the VRT system; identifying whether the system will be guided by map-based inputs or sensor-based inputs; and the identification of the kind of data that will be used for mapping or the kind of data that will be sensed for actuation of the dispensers. In terms of the kinds of inputs, sensor-based inputs are more sophisticated because they are continuously updated based on the changing conditions on the farm. At the same time, they are also more logistically and computationally expensive because they mean the battery-powered sensor nodes will need to continuously or intermittently (guided by algorithmic inputs related to outlier or bottleneck detection such as in~\cite{RL-2018}) relay data to the actuators for VRT dispensing. For farmers to adopt these technologies, concrete savings on fertilizers need to be demonstrated with potential yield increase and environmental protection from decreased farm effluents from nutrient pollution, reducing farm runoff and eutrophication (hypertrophication), such as from high levels of nitrogen and phosphorous in fresh water. In the case of livestock farmers, this translates to the decreased use of hormones or antibiotics for the livestock.
%https://www.epa.gov/nutrientpollution/sources-and-solutions-agriculture

There  are  important  questions  to  be  answered  before  we  can  leap  into  data  analytics  for  agriculture,  questions  related  to technical viability, economic feasibility, sustainability, and data protection and ownership. These questions cannot be looked at in isolation---for  example,  if  some  algorithm  needs  data  from  multiple  data  owners  to  be  pooled  together,  that  raises  the  question  of  data ownership and data privacy. Data privacy is especially important when the raw data and the algorithm needs to be fed to cloud-based Machine learning-as-a-service (MLaaS) computing platforms~\cite{differentially-2019} or to collaborative learning based on training data fed by multiple edge devices~\cite{obfuscate-2018}. Here, data perturbation and encryption needs to be orchestrated without significantly affecting model outputs.  This  paper  is  the  first  to  bring  together  these  questions  under  one  umbrella, discussing the goals, path forward, and challenges of digital agriculture to surmount the nexus-sensitive challenges.
\section{Data Generation from Sensors}
 This section will cover the modalities of data gathering and controlled dissemination, the volume of data generated, and the quality of the data, collected and processed from the large numbers of inexpensive sensors proliferating in farms.
Fields, these days, are increasingly equipped with sensors for sensing the multi-dimensional attributes that determine the quality of the agricultural field, mapped to the resultant crop productivity. Here are some of the commonly used measurements from agricultural sensors.

\begin{itemize}
    \item {\textbf{Soil sampling}:} Soil cores are extracted from field (may or may not be georeferenced) and sent to a lab for analysis. Lab results come back in a report detailing fertility levels. In most cases, one must manually assign geo-referenced points to report values.
    \item {\textbf{Fertilizer application}:} Based on the results of the soil sampling, a fertilizer recommendation is generated. If geo-referenced points are used, a variable rate prescription can be generated. If fertility results are aggregated across the field, a flat rate is applied, maybe proprietary or shapefile.
    \item {\textbf{Planting}:} Can generate as applied maps that include information on population, singlulation, misses, again  proprietary or shapefile.
    \item {\textbf{Scouting}:} Conducted as frequently as once a week during growing season. Traditionally data comes in from of a report that details presence of disease, insect, and weed pressure. Currently much research is being conducted on drone based scouting using multi-spectral imagery to determine nutrient and water deficiencies as well as detect disease, insect, and weed pressure. Data format is large image files that need post processing.
    \item {\textbf{Spraying}:} Based on the results of the scouting reports spaying operations are conducted. Most modern sprayers can generate as applied maps. Files are saved in a proprietary format based on the sprayer manufacturer.
    \item {\textbf{Harvesting}:} Yield maps are generated by harvesters and saved in a proprietary format based on the manufacturer. In irrigated fields, soil moisture sensors are often used to determine irrigation intervals. Despite the wealth of available data in most operations, very little is actual analysis and used to inform future decisions. The most commonly used data sets are scouting data (used to make chemical application decisions) and soil sampling data (used to make fertilizer recommendations). More progressive producers use yield data to determine aid in the generation of variable rate fertilizer application and for comparison of seed varieties. Much of the data collected only contain value when aggregated with additional local or regional data. 
    \end{itemize}

In terms of VRT, which coincides with advancements in electronic controls and improved communication technologies, the following could be applications of the technology: fertilizer application both macro- and micro- nutrients, herbicidal and pesticidal applications, manure, seed applications, tillage as a function of soil compaction, and precise irrigation. Thus, if the VRT procedures are guided by sensors rather than more static maps, these could be some of the farm processes benefiting from the approach.

\textit{Aggregation of agricultural data is possibly the biggest challenge facing digital agriculture.} Producers are reluctant to share data due to fears of regulatory issues and the lack of perceived value added to their operation. Until producers can clearly see the economic benefits of digital agriculture, adoption will be slow. 
%\newpage

\section{Data Lifecycle}
This section will cover the various phases in the lifecycle of big agricultural data---sanitization, loading, processing, storing, summarization, and analysis; an example is shown in \textbf{{Figure~\ref{fig:ag-data}}}. This will go into some of the general-purpose approaches (such as, data deduplication, calibration using sensor metadata) as well as agricultural-specific approaches (such as, known variations in hyperspectral maps from ground sensing and aerial image data and effective fusion among sensor arrays.
\subsection{Data generation sources}
\noindent Data generation is the first stage of a data lifecycle. There are many ways in which data can be generated. The sources of data generation can be broadly classified into two types. 
\begin{enumerate}
    \item \emph{\textbf{Localized data or private data}:} This is the data that is generated on the farm such as soil nutrient composition, water, and fertilizer usage. This type of data is generated from sensors that are present on the farm.
    \item \emph{\textbf{Public data}:} Data such as historic weather conditions and market prices fall under this category. Imported data is often generated at outside sources and shared with the farmers to use in precision agriculture. Such data is not farm specific. An example of data that is at the crux of localized and public data is topography and soil type, which may be somewhat localized but follow a trend for farms in geographical proximity.
\end{enumerate}

\subsection{Data warehousing}
\noindent Data generated then needs to be stored in repositories called data warehouses. Data warehousing allows integration of different data from multiple sources and helps restructure the data for better performance. One recent example of data warehousing is an initiative taken from the government of India~\cite{sharma2000integrated}, titled INARIS (\emph{Integrated National Agricultural Resources Information System}). \cite{woodard2016big} discusses Ag-Analytics---a platform that provides data warehousing in the field of precision agriculture.  Although there are readily available platforms for data warehousing, there are several constraints when using these platforms directly in precision agriculture. Some of these constraints are discussed in \cite{nilakanta2008dimensional}.

\subsection{Data annotation and cleaning}
\noindent Due to the large size of data, it is important to perform annotation and cleaning before data analysis. Data annotation is subjective and depends on the particular use case of precision agriculture. The choice of the data annotation technique is dictated by the size of the data set, cost of annotation per sample among many other guidelines. \cite{schoofs2010annot} proposes a data annotation technique for electricity data in wireless sensor networks (WSNs). Similar annotation techniques could be developed in the aspect of precision agriculture that can be performed in a WSN infrastructure. 

Data cleaning removes or corrects errors that are present in the data. There are several existing works that propose data cleaning techniques in precision agriculture.
% http://www.gisagmaps.com/yield-monitor-data-cleaning/; K. Sudduth in the yield data cleaning
\cite{simbahan2004screening} proposed a screening algorithm for cleaning yield data that provided an increase in map precision. \cite{sun2013integrated} proposes an integrated framework for software that increases mean yield through data cleaning.

\subsection{Metadata annotation}
\noindent Metadata annotation can be done manually or be automated. Since the data that we are considering is prone to be complex, automated metadata annotation is preferred. \cite{roy2010comparative} provides a comparative study of the different learning techniques used for metadata annotation. \cite{fiehn2005setup} provides an algorithm for annotation of metadata.  \cite{haug2014crop} uses a human expert to first mark crops from raw images. Masks were then derived from using these markings. These masks are then used to acquire the metadata. Similar techniques can be used in different aspects of precision agriculture.

\subsection{Data processing}
\noindent The last stage in the lifecycle of data. \cite{steven1993satellite} provides a good overview of the constraints faced in data processing in precision agriculture. In~\cite{steven1993satellite} the authors consider the case of using satellite images for remote sensing in precision algorithms. In such use cases, one of the important aspects of data processing must be to make the data more readable. One such scenario where these images can not be directly used is in the case of cloud cover, where the images need to be processed before utilizing the data. This can be extended to data acquired through other means as well. Data acquired from soil sensors may need to be processed in order to make it more utilizable. \cite{honkavaara2013processing} proposes a processing chain that uses data collected from unmanned airborne vehicles to generate meaningful results. \cite{loreto1996development} proposes an automated system that performs both the data acquisition and data processing of soil nitrate measurements. \cite{murakami2007infrastructure} proposes a data processing algorithm that processes yield data on a distributed systems framework. 
\section{ML and Low-Power Communication Technologies}
Here, we will cover approximate processing for in-sensor analytics, advanced processing for backend analytics on the edge or the cloud platforms, and interpretable data analytics. The first aspect is particularly useful because of the expensive nature of the wireless communication or the lack of continuous connectivity of the sensors to the backend. The second aspect is useful because there are edge platforms for agricultural data, such as, Azure IoT Edge device~\cite{farmbeats-2017}, and there may be sensitivity of the farmer to upload her personal data to a cloud platform. The final aspect is important because the farmer, a non-data science expert, will still need to be given some insights into the results of the algorithm, at her level of understanding, prompting her to take actionable measures.

Data analytics plays an important role in precision agriculture. It can help farmers decide what crop to grow when, monitor the crop growth, and decide on the logistics of farm management. But agricultural data is often large and noisy and needs careful processing to distill insights from them. The following subsections elaborate on the advanced ML capabilities that can be used for such analysis. In addition, another relevant technology in this context is the use of scalable databases to house and process these data sets for downstream processing and retrieval.
% , including the use of ML for actionable insights for farmers and for agro-entrepreneurs. 
With this in mind, we also include some innovations in NoSQL database technologies to assist in high-throughput information retrieval from the evolving agricultural data lakes. With the advent of more sophistication in farm machinery and nutrient application, another related aspect is the reduction in latency during the application of agronutrients. In such cases, the use of low-latency channels, such as the deployment of intelligent in-sensor processing and low-latency edge processing will also be discussed.

\subsection{Approximate processing for in-sensor analytics}
\noindent A sensor network acquires real-world measurements at discrete points, where each measurement is a snapshot in time and space. In most scenarios in which sensor networks are deployed, the sensors are frequently queried resulting in continuously monitoring alongside high energy costs. 
Thus, one of the chief problems faced by wireless sensor networks (WSN), often deployed in farm settings, is the constrained availability of resources to these devices. The sensors used in such 
networks are low-power embedded devices that are expected to last for long periods of time (order of months) on standard batteries~\cite{raza2017low}. This issue can be mitigated by leveraging a more distributed architecture and using more energy-efficient algorithms. Further, these devices generate large volumes of data, which ideally will be processed in real time in a streaming manner for usable insights. 
% Traditional analytics would not work well in such settings because of the huge draining of computational resources from these continuously-ON devices.
% \emph{Internet of Things} (IoT) devices are deployed in the environment to sense or 
% collect data. It is predicted that the data generated by such devices would reach 
% 508 ZB by 2019 \cite{sanyal2018improving}. 
As a tradeoff between computational load and accuracy, approximate computing tools and techniques can be deployed~\cite{mitra2017phase}.
The idea is to perform approximate computation over carefully chosen subset of the entire input data set. For example, the approximation can relay whether a particular soil nutrient concentration is above or below a threshold, rather than the exact value of it. Also, it may compute this over a uniform random subsample of say one in every 10 samples. Alternately, we can also use information theory principles, such as the Nyquist-Shannon sampling theorem to decide on the spacing of the sensors in the WSN~\cite{qin2015wideband} and also the actual redundancy needed for robust sampling from these sensors~\cite{shin2013optimization}.

An important requirement of our target applications is low latency. This can be achieved by using multiple nodes to parallelize the work. However, this has to be done carefully so that the load is approximately balanced and there is not much overhead of energy to perform the distribution. For example, ApproxIOT~\cite{wen2018approxiot} proposes an algorithm based on Apache Kafka~\cite{thein2014apache} that uses IoT devices to generate data and forward it to edge computers managed by service providers. These data streams are then sampled and forwarded to a central location, where user-specific queries can be 
made. The sampling in such systems are based on two techniques, 
\begin{itemize}
    \item \textit{Stratified Sampling:} The streams of data are categorized based on their distribution. A random sampling is done on these distributions, with prior knowledge of the data required for this kind of sampling.
    \item \textit{Reservoir Sampling:} A reservoir size $R$ is maintained and at most $R$ items are uniformly sampled from the data set. Here, prior knowledge of the data is not required.
\end{itemize}
ApproxIOT extends both these techniques to propose a weighted hierarchical sampling. The nodes conduct sampling over data generated and compute statistics, resulting in a 1.3X to 10X speedup.

In-sensor analytics is relevant to our domain because it means that the data being sensed will be (partially) analyzed locally at the sensor itself. The value proposition is that raw data will not have to be sent over the wireless network, thus saving on wireless bandwidth and energy, and thus potentially yielding low-latency decisions. 
In-sensor analytics algorithms can either be \textbf{value tolerant}, where the 
approximation of values can be made and the resulting algorithms are lightweight 
or they can be \textbf{delay tolerant}, where the resulting algorithms can be more accurate at the expense of latency.

SERENE~\cite{baralis2006selecting} is a framework that selects \emph{representative} 
nodes, among clusters of correlated sensors. Here, the framework also takes into account the dynamic changes in the network topology and the presence of outliers. Since sensor data acquisition and communication are the main power guzzlers and sensors are typically battery-powered, SERENE uses clusering algorithms to spatially and temporally aggregate the data. For example, it uses a density-based clustering algorithm, DBSCAN~\cite{ester1996density} for robustness against outliers and noise. This algorithm can cluster based on any shape, as the sensor readings maybe correlated. 
Based on the cluster shape, availability of battery power, and distance of the target node from other nodes, the representative nodes---\emph{$M$-sensors}---are queried. %When the user needs to query some information, only these $M$-sensors will be queried. 
%Since these $M$-sensors are representative of the entire network, an approximate analysis can be made on the information returned. 

Another lightweight approach for in-sensor analytics is Snapshot Queries~\cite{kotidis2005snapshot}. Here the representative nodes are elected through a localized process. Each node maintains a data distribution model of its neighboring nodes. Based on this model, it predicts the values of the neighboring nodes. If the error between the predicted value and the actual value is less than a threshold, the predicting node can represent the neighboring node in question. The data model maintained is based on the previous correlation between the values of the node and its neighbors. Here the model is not static and is revised frequently. Hence it is robust to dynamic changes in the network. 

Kartakis \emph{et al.}~\cite{kartakis2014real} proposed a scheme that reduces the usage of computational resources by 85\% and memory resources by 55\%. The main motivation for this work was that there is a correlation between the compression rates and data fluctuation. Hence the compression rate can be used to detect anomalies and outliers. Kalman filters are used to reduce the number of false positives. The Kalman state is updated by feeding every new input to the filter. The anomalies are detected as huge drops in the Kalman states. 

SnowFort~\cite{liao2014snowfort} is another open-source WSN that is energy-efficient and uses open-source hardware in concert with a new open-source communication scheme to allow for reliable infrastructure and environmental monitoring. The architecture used in Snowfort is a three-tiered architecture, with wireless sensor nodes, a base station, and a cloud server. 
% The WSN is the sensing unit that aggregates data. 
% http://web.stanford.edu/group/snowfort/
The base station or access point intermediates a group of nodes to form a network. The cloud server is the decision making unit. This type of decentralization reduces the load on the sensor nodes.

The concept of lightweight algorithms can also be extended to deep neural networks. Distributed Deep Neural Networks (DDNNs) provide better fault tolerance and security than DNNs. The data generated by sensor nodes is processed locally at the edge. %As would be expected, a face-recognition algorithm has a reduced reflex time when the photos are processed at the edge rather than on the phone~\cite{yi2015fog}. 
DDNNs are used to utilize the advantages of distributed computing hierarchy in DNNs~\cite{teerapittayanon2017distributed}. 

Another aspect to reduce the energy consumption of sensor nodes is by \textit{reducing the duty cycles (sleep-wake cycling) of sensors}. By activating the sensors only when required, the energy consumed by the sensors will be reduced. One way to achieve this is by using time multiplexing like in WirelessHART~\cite{chen2010wirelesshart}. This also provides an added advantage of non-interference. Mazo and Tabuada~\cite{mazo2011decentralized} consider a decentralized non-linear control system with a feedback control law. They continuously check certain parameters and if a certain relation between these parameters is satisfied, the computational law is recomputed providing an event-triggered implementation. 

\textit{Example of trade-off between computational accuracy and resource usage:}
Sensors used in precision agriculture are \emph{Internet of Things} or IoT devices and like all IoT devices have constraints on power and connectivity. It is necessary to have the first step of data processing that takes place at the sensor's end to be energy efficient. Thus, anomalous data can be suppressed. Alternately, in some scenarios, the exact opposite is desired---when an anomalous event is detected, that event needs to be communicated to the gateway.

AutoRegressive Integrated Moving Average (ARIMA) is used for fitting time-series data in order to predict or estimate the trend. This is done by using regression on the system's prior values.  Autoregressive models find a linear relationship between current values and prior values of the  system, while moving-average models find a relationship between the current value using previous and current residuals. ARIMA models combine both these techniques to transform a non-stationary time series to a stationary time series using a differencing approach. Differencing stabilizes the mean of the time series by reducing outliers in the time series. 
Exponential Smoothing (ETS) techniques are used for smoothed curve fitting using exponential time windows.
ETS and moving average models both introduce a lag with respect to the input data. They also have the same distribution of forecast errors. ETS takes into account all the past values and current value but only requires the most recent value to be computed, while moving average models only take into account the previous $k$ values and the current value and requires $k+1$ data points.

While ARIMA models work on a linear process, more sophisticated ML algorithms can model non-linear processes. While traditional ML techniques such as \emph{$k$-Nearest Neighbors} (KNNs) and \emph{Support Vector Regression} (SVR) have more complexity than ARIMA or ETS, they do not demonstrate better performance than ARIMA models. \emph{Recurring Neural Networks} (RNNs) \cite{elman1990finding} and \emph{Long Short Term Memory Neural Networks} (LSTMs) \cite{hochreiter1997long}, however, provide a considerable tradeoff. RNN or Elman network contains feedback connections that provide a relationship between the previous states and the current input in order to produce a final output. By preserving copies of previous values of recurrent layers, RNNs encapsulate historical events in the modeling. Hence, dynamic temporal behavior for a time sequence is observed. LSTMs have a similar working principle as that of RNNs, albeit they preserve information over longer time periods. 

While RNNs and LSTMs give a superior performance over statistical methods like ARIMA and ETS, they are more complex in nature. For simpler user queries and in-sensor analytics, ARIMA and ETS models can be used as a lightweight alternative. A more detailed analysis can be done on the edge (or cloud platform) using more sophisticated ML approaches of the likes of RNN and LSTM models or with some context imparted through \textit{attention pooling} for example. Makridakis \emph{et al.}~\cite{makridakis2018statistical} a detailed comparison between the performance, model fitting, and complexity of different ML techniques. 

\begin{comment}
\begin{wrapfigure}{l}{0.44\textwidth}
\begin{center}
\includegraphics[width=\linewidth]{complexity.png}
\end{center}
\figurecaptionspace
\caption{A comparison of performance error with computational complexity of various statistical and ML methods \cite{makridakis2018statistical}.}
\label{fig:ML-complexity}
\end{wrapfigure}
\end{comment}

\subsection{Database management and backend analytics for large-scale agricultural sensor data}
Precision agriculture allows for site-specific crop management to increase throughput and achieve more sustainable farming by applying data science to agriculture practices, learning from local data trends. More and more agricultural sensing data is being live-streamed from farms, whether it be through on-board cameras, on manned or unmanned aerial vehicles, or through ground sensors in the farms. There is thus a need for centralized databases to store and process these data sets, often in real-time, to get actionable insights for farmers. Plus, there may be some degree of federation in storage and compute resources that may be needed as the computing needs of this domain increases, as has been seen in the genomics domain~\cite{federation-chaterji2019}.
For example, if decision can be made about the level of application of some fertilizer while the dispenser is moving through the field, this will be advantageous. 
Further, the sensing data is multi-dimensional and noisy, coming from ground sensors deployed in farms to measure an array of soil characteristics, such as, moisture, nutrient levels, temperature profiles, soil acidity, etc. These live-streamed data sets need to be stored in a fail-safe repository of nodes, such as Redis installations in Amazon Web Services (AWS) Elastic Compute Cloud (EC2)~\cite{carlson2013redis}. The aggregate live-streamed updates and queries represent a unique workload for such NoSQL datastores, as these queries vary in rate and type over time. In such cases, a workload-aware tuning system is needed to reconfigure the NoSQL cluster, whether locally or on the cloud, to provide high performance. This is becoming more important as the data from small-scale farms across the country is burgeoning both in size and diversity, slowly replacing the previously used manual data-collection processes. 
Maximizing the throughput of these processing pipelines of digital farm data will enable actionable insights from agricultural sensing data.  Also, given that these pipelines are hosted on the cloud, we want to maximize the performance within a user-defined cost bound, as shown in our recent work on cost-aware optimization of noSQL database throughputs~\cite{sophia-atc2019, rafiki-middleware2017}. A  pipeline for analytical workloads (OLAP, online analytical processing) consists of two main parts: a storage cluster (\eg Redis or Cassandra) and a computing cluster (\eg Spark), the latter operates on the top of the storage cluster to train ML models and execute complex analysis on the stored data.
The task of the optimization is to find the best combination of configurations maximizing the objective metric, modeled as follows: 

\begin{equation}
    P^* = \argmax_{Confs} f(Confs(t), WL(t)) 
    \label{eq:Optimizer}
\end{equation}

Where $P^*$ is the optimal performance that is achieved by the best combination of configurations $Confs$. The search space of $Confs$ is large, \eg noSQL databases, such as Cassandra and Redis, have 50+ and 40+ performance-sensitive configuration parameters, respectively. Hence, an exhaustive search through all possible configurations is impractical. Therefore, evolutionary search techniques are preferred in this case due to their ability to find close-to-optimal solutions in practical time, \eg \textbf{Equation \ref{eq:Optimizer}}. 

\subsection{Microservices and edge-cloud partitioning for low-latency communications}
\noindent The world of connected devices has fueled the IoT era, where applications rely on a multitude of devices aggregating and processing data sets across highly heterogeneous networks. In this context, distributed deployment alongside containerization of the different information channels will shield the systems from isolated failures, conferring resiliency. The other important aspect is the partitioning of the data stream for computing at different degrees of latency---computing nodes at the edge are used for user-facing applications (emails, order placements, payroll, etc.) and the owners will react negatively if the computers become unusable due to intermittent edge analytics. Therefore, the prioritization of the processes need to change dynamically. In contrast, the application itself needs to be designed in a way that it is insensitive to such dynamic, and unpredictable, changes to the priority level, e.g., it will not time out if there are client-server interactions. Another aspect of the prioritization is that the different analytics results are needed with vastly differing timing requirements. Such high-level, user-expressed requirements will be used to dynamically prioritize in the face of unpredictable arrivals of the events (e.g., a flash flood event, or onset of a locust infestation). Thus, overall the partitioning needs to happen in a top-down or bottom-up manner. Top-down means that we take the high-level user requirements on latency and accuracy and define the partitioning based on that. Bottom-up means that depending on the available resources on each platform, the resource handler decides where to run the application component. Top-down requirements naturally have a higher priority. This will leverage the significant amount of work that has been done in automatic partitioning of applications to run on mobile devices and the cloud~\cite{kovachev2012adaptive, mora2017distributed, singh2017optimize, li2018auto, dey2018partitioning, wang2019edge}.

\subsection{Low-power communication technologies}
\noindent Low-power communication technologies for wireless IoT communication falls broadly in three categories: 

\begin{itemize}
  \item Low-power wide area networks (LPWAN), with a greater than 1 kilometer range, essentially low-power versions of cellular networks, with each ``node'' covering thousands of end devices. Examples include LoRaWAN, Sigfox, DASH7, and weightless.

  \item Wireless personal area networks (WPAN), typically ranging from 10 to a few 100 meters. Examples include Bluetooth and Bluetooth Low Energy (BLE), ANT, and ZigBee, which are applicable directly in short-range personal area networks or if organized as mesh networks and with higher transmit power, larger coverage areas.
  
  \item Cellular solution of IoT, including any protocol that are reliant on the cellular connection
\end{itemize}

Some of the bottlenecks in wireless transmission in farm settings include the harsh physical conditions in farm settings and the proliferation of inexpensive and less reliable sensors coupled with the intrinsic challenges of the LoRaWAN and cellular network technologies, such as 5G.

\subsection{Interpretable data analytics}
\noindent Although there are several predictive analyses models, it is important that the analysis obtained be interpretable. \cite{molnar2018interpretable} explains the importance of interpretable ML. This could help agriculturalists better understand the crop growth and manage it better. \cite{vellido2012making} provides an overview of several works that address the issue of making ML techniques more interpretable. Dimensionality reduction is a popular choice in this aspect. A comparison of such existing techniques is provided in~\cite{van2009dimensionality}.
After dimensionality reduction, it becomes tractable to rank order the different features by their importance, thus providing an important insight to users of the models~\cite{interpretable-chaterji}. Another aspect of interpretability is to answer post-hoc ``sanity check'' questions~\cite{inouye2018deep}, such as, after a rain event, does the model predict the moisture content in the soil is higher. If the model fails to answer correctly a sanity-check question, then further examination of the model is done. 
%how does dimensionality reduction enable interpretability

\section{Democratizing AI for Farmers}
\noindent This will require the accessibility of state-of-the-art wireless, database, and ML technologies to farms---even small-scale farms---deploying leading-edge analytics and broadband modalities to gather data from farms (\eg FarmBeats, AirBand technologies, Open Ag Data Alliance APIs). This will cover the commercial angle of deploying and operating a system and the foundational (and often open-source) technologies that they build upon, such as, TV whitespace spectrum being used to upload high volume data.
To bring the benefits of digital agriculture, and the ML driven
insights to the farmers, we need to make these systems
affordable. However, existing digital agriculture solutions are
expensive for two key reasons. 

\underline{First}, is there Internet access, both in the farmer's house, as well as on
the farm. The sensors, drones, and tractors need to send data to the
cloud to train the ML models, and the insights need to be conveyed to
the farmer's devices. However, the farms are far remote, and do not
have access to affordable, broadband Internet. In fact, nearly half the
population of the world, most of whom live is rural areas, do not have
Internet access. Even in the US, around 20 million rural Americans
do not have broadband access~\cite{Broadband}. And a large part of
farmland does not have access either.
%https://www.fcc.gov/reports-research/reports/broadband-progress-reports/eighth-broadband-progress-report
To bridge this digital divide, we need to come up with innovative
solutions to bring broadband Internet in the farms. One such promising
technology is the TV White Spaces, which refers to unused TV
channels~\cite{white2009}. There is abundant unused TV channels in
farms, which can be used to send and receive data in farms. Wireless
signals also propagate smoothly in the TV frequencies, and are ideal for connecting farms. Satellite-based Internet access also holds promise, where low-earth orbit (LEO) satellites could
enable low-cost backhaul Internet access.

\underline{Second}, the reason for expensive digital agriculture solutions is the
lack of platforms of innovation in agriculture. Data acquisition
systems are proprietary, and hence getting data from the farm for
research is not easy. There is also a dearth of clean data from farms
that can be used for building ML models. 
We propose the building of open-source APIs, and data
repositories. These APIs, such as the one from Open Ag Data Alliance,
afford researchers and startups a platform to prototype their
innovation. Similarly, a cloud-based data repository, for example with
drone imagery and satellite data from research farms, will enable
researchers to train new models. It will also help create a
benchmarking data set to evaluate new innovations in agriculture before
bringing it to the growers. 

\section{Big Data Frameworks for Agricultural Data}
\noindent Here we will discuss the popular frameworks through which we will invoke the ML algorithms. This will involve open-source frameworks such as Apache Spark, streaming data processing frameworks such as Apache Flink, and techniques for distributing the ML processing among nodes in a cluster.
\subsection{Streaming data processing}
\noindent Since most of the data obtained in digital agriculture is real time, stream processing is preferred over batch processing. This adds several constraints: {\em first}, the data analytics code has to function at a rate at least as fast as the rate at which data is being generated; {\em second}, it has to calculate statistics (such as, range for normalization of data) without access to the entire data and based on some look-ahead window based on the workload dynamism (for example, for a more dynamic workload, the look-ahead window will be lower for higher accuracy, plus the algorithm should be configured for some degree of error handling); and {\em third}, the code has to have the right input-output interfaces so that it can ingest streaming data and output its results in a stream. Further, Now we consider some popular open source streaming analytics frameworks---Apache Spark Streaming (\url{https://spark.apache.org/streaming/}), Apache Storm (\url{http://storm.apache.org/}), Apache Flink (\url{https://flink.apache.org/}). These differ in the ways in which they can transform the data stream, i.e., the kinds of operators that they support, the latency of processing, the programming languages they support, etc. 
\cite{carbone2015apache} provides an open-source stream processing framework. Flink however does not provide its own  data storage and needs to be supplemented by frameworks that do, such as Kafka~\cite{garg2013apache} or Cassandra~\cite{cassandra2014apache}. Apache Spark~\cite{zaharia2016apache} is an alternative that can also be used for data parallelism. 
%talk about data and model parallelism here

\subsection{Batch data processing}
There are some data analytics applications that need batch processing in this domain. This includes typically analytics that will be processed for strategic decision-making, which does not have any real-time requirement. Batch data processing is done through data warehousing tools~\cite{chaudhuri1997overview, nargesian2019data} and analytics frameworks like Spark (as opposed to Spark-Streaming) that can ingest data from such warehouses. The ease of this mode of processing is that there are no real-time requirements and the analytics code can access the entire data in one shot. The challenge with this mode of processing is the large volume of data. To fit within the resources of the compute nodes, the data has to be segmented and the analytics code in practice runs on the segmented data.

\subsection{Open-source frameworks}
\noindent Open-source frameworks provide extensive customization and allow collaboration. These two properties among many make them more preferable. \cite{hashem2015rise} provides an overview of the existing challenges and solutions to handling big data. Hadoop is one such open-source framework that can process large data sets \cite{shvachko2010hadoop}. It functions on map-reduce programming models. Several warehousing solutions built on Hadoop such as Hive~\cite{thusoo2009hive} are also available. Apache Drill is another open-source framework that provides interactive analysis of big data~\cite{hausenblas2013apache}. \cite{chandarana2014big} provides a comparative study of three open-source frameworks, Hadoop, Drill, and the Project Storm.

\section{Analytics for Alternative Agriculture }
\noindent Increasing global energy demand and environmental concerns associated with petroleum have raised interest in carbon-neutral biofuels for reducing dependency on fossil fuels that result in human-caused (anthropogenic) greenhouse gases. Among biofuels, although ethanol is a renewable biofuel in use, made from biomass feedstocks, it contains about 30\% lesser energy than gasoline per gallon. Advanced biofuels like butanol, isobutanol, fatty-acid and isoprenoid-derivatives are more energy dense with combustion properties similar to existing fuels. However, the native pathway of microbes is significantly low for these fuels for commercialization. Metabolic engineering using control theory and ML approaches can redirect the cellular fluxes toward improving the titer of the microbial synthesis of these higher energy density biofuels. 
 advanced biofuels. Advanced biofuels are less volatile than ethanol and are often produced from lignocellulose destruction, the structural framework for which is in the ballpark of 30-50\% cellulose, 15-35\% hemicelluose, and 10-20\% lignin~\cite{leitner2017advanced}. The biofuel is generated by first breaking down the starting materials into small ingredients, followed by hydrolysis to fermentable sugar products, which are finally fermented to biofuels. This is in line with thinking of the entire managed ecosystem of ``farming'', whether it be for food crops, biofuels, or fibers. Following are some of the high-level optimization strategies that can be leveraged for advanced biofuel production, an exemplar alternative agriculture example for this vision article.
\paragraph{Microbial production of advanced biofuels:} 
Microorganisms can be engineered to produce bulk chemicals such as biofuels but the produced chemicals are often toxic to the cultured cells, and there is always a trade-off between the biofuel production and cell survival, limiting biofuel production. One of the possible ways to increase biofuel production is to increase the tolerance of the cultured cells toward the end product chemical. Biofuel tolerance can be improved by reprogramming the innate metabolism of these microorganisms, such as by using genome engineering~\cite{chaterji2017crispr, PAM2019} or the newer prokaryotic genome editors, which do not require the protospacer adjacent motif (PAM) sequence for cleavage~\cite{fu2019prokaryotic}. However, tolerance to biofuels like ethanol, or glucose mixtures is not a monogenetic trait. It requires mutations in multiple genes and pathways to be accurately mapped. There are several ways in which this mapping can be ``learned'', one of them is to culture the cells in a stressful environment containing high concentration of the chemical, for example, ethanol, against which the tolerance of the cells, for example, \textit{E. coli}, is desired to be improved. This triggers mutations in the cells as a response to adapt in the given environment of high ethanol stress. These mutations could be causal, that is, adaptive mutations triggered by the high stress, or non-causal passenger mutations. Identifying the adaptive mutations from passenger mutations is a difficult problem. Studies have successfully identified single genes that are responsible for the adaptive mutation toward biofuel tolerance, albeit tolerance is a complex trait influenced by multiple genes and pathways. 
Near-lethal stress conditions can enable evolution producing populations that have acquired a hypermutator phenotype and are tolerant toward the stress applied. The genes and pathways responsible for this adaptive mutation  can be identified using network-based computational approaches. This requires analyzing the temporal profile of adaptation to ethanol stress. The identification of mutations occurring during the evolution of the microorganism, coupled with the functional implication of each mutation at the protein level can provide us with enough information to computationally map out the adaptive pathways that lead to increased tolerance toward a certain biofuel like ethanol. Once the results are found to be consistent in independent populations of a certain microorganism, the hypothesis can be validated in actual laboratory settings by genetically modifying the organism's genome. 
For example, the genome sequencing data of \textit{E. coli} (strain SX4) that gradually evolved to tolerate high ethanol stress is available in the SRA repository of NCBI, PRJNA380734 (https://www.ncbi.nlm.nih.gov/bioproject/380734). The data for yeast evolving under ethanol stress is available as the YEASTRACT database (http://www.yeastract.com/). The mutations in the genes could be identified over time, and the corresponding genes could be maintained as nodes in a gene interaction graph. Such a data structure could make it easy to identify the pathways that lead to increased tolerance toward ethanol stress. Not all mutations are likely to be a part of the adaptive phenotype, and therefore, in previous studies, for example~\cite{Swings}, a relevance score was computed for all mutations based on some prior information and the frequency of occurrence during a population's increased fitness. From this graph, a sub-graph has to be computed in a way that genes with large relevance scores are chosen and the paths within the sub-graph have a higher frequency of occurrence. Such computation needs to take care that the results are least affected by noise due to passenger mutations. Since the overall approach is mathematical, the computations may be repeated with varying parameters that will result in different sized sub-graph identification, or selection of mutated genes above a certain relevance score or pathways with occurrence frequency above a certain threshold. After complete analysis of the results, the best results can be chosen based on certain criteria that reflect actual biological phenomena, and biological validation can be done.
   
\section{Economics, Policy, and Decision Making}
\noindent We will discuss the policy issues that should regulate the use of big data and the economic factors that will be important for the adoption of big data. The most relevant policy questions in digital agriculture regard the value and legal status of farm and related business data. Though farm data enjoy some of the intellectual property protections afforded to trade secrets, its legal ownership structure remains ambiguous~\cite{miller2018estimating, ferrell2016legal}. The decision to subscribe to a data service provider may be impacted by fears of personally identifiable information (PII) being misappropriated. Yet farm data generates positive network externalities when aggregated across a large number of operations. Business models, such as Farmers Business Network, Inc. (FBN), have demonstrated the value of data sharing. Academic research should focus on designing incentive-compatible mechanisms that encourage data sharing with public universities while protecting farmers’ intellectual property (IP).

\subsection{Profitability and on-farm decision making}
\noindent Site-specific farm management has the potential to enhance farm profitability while conserving resources. But despite advancements in field sensors, GPS guidance, and grid soil sampling, adoption by farmer operators of has fallen short of expectations \cite{schimmelpfennig2016farm, bullock2009value, baerenklau2007dynamics}. Operator demographics, operation size, and perceived benefits influence the decision to invest in site-specific management practices \cite{thompson2019farmer, schimmelpfennig2016farm}. Subsequent economic returns to adoption depend on the nature of the adopted technology and its interface with on-farm decision making \cite{bullock2000agronomic}. 

Miller \textit{et al.} (2019)~\cite{miller2019bundle} identify two types of precision agriculture technologies: embodied knowledge---tools that generate value in isolation such as GPS guidance systems or automatic section control---and information intensive---tools that produce data for use in future decision making such as yield monitors, grid soil sampling, or electro-chemical sensors. Embodied knowledge technologies create convenience the moment they are employed while the benefits of information-intensive technologies are revealed over a longer time horizon and depend on their role in the on-farm decision making process~\cite{thompson2019farmer}. Differences in the immediacy and measurability of realized gains may explain differences in adoption rates across technology types. Use of variable rate technology (VRT) and GPS soil mapping, for example, has consistently lagged that of GPS guidance systems~\cite{schimmelpfennig2016farm}.

Rather than assessing technologies in isolation, agricultural economists are increasing interested in how producers bundle complementary tools to create an overall precision technology strategy~\cite{schimmelpfennig2016sequential, miller2019bundle, lambert2015cotton}. Profitable use of ``hard” technologies'' such as variable rate planters and fertilizer spreaders depends crucially on the availability of accurate intra-field soil data, or ``soft'' technology inputs~\cite{bullock2009value}. These data sources however, are themselves costly to obtain. Moreover, the optimal data collection frequency or sampling density is not obvious and likely varies by field~\cite{franzen1995sampling}. 

Unless the economic returns to site-specific management cover both the up-front investment and the cost of collecting actionable information, adoption will be low. The unit cost of grid soil sampling for example---traditionally around \$10 per sample performed by manually extracting cores and sending to a lab for analysis---will fall as automated sampling technologies become commercially viable. This does not however, guarantee a profitable result from site-specific farm management. Without a sufficient amount of within field variation, uniform input application may be economically optimal. Future research on decision making in the field of digital agriculture should focus both on the adoption decision and how producers optimally implement technologies given their information constraints.

\subsection{Policy}
\noindent The most relevant policy questions in digital agriculture regard the value and legal status of farm data. Agricultural data generates benefits when aggregated across a large number of farm operations~\cite{miller2018estimating}. These benefits---referred to as network effects or network externalities---grow with the number of participants~\cite{rohlfs1974network}. Business models such as Farmers Business Network, Inc. (FBN) have demonstrated the value of data sharing through its crowd-sourced database of input costs and performance bench-marking. But farm data often remains siloed within the farm gate~\cite{coble2018bigdata}. An individual farm's data leads to more reliable recommendations when pooled with comparable operations using similar practices and inputs. To overcome this ``small data'' problem, the perceived benefits of joining a big data community must exceed the perceived costs---most of which stem from privacy concerns over data misappropriation~\cite{griffin2016property}. In particular, farm operators may fear that personally identifiable information (PII) could be used against them by regulators or environmental activists~\cite{ferrell2016legal}.

To better understand the privacy concerns of producers, the nature and legal status of agricultural data must be considered. Miller et al. (2018)~\cite{miller2018estimating} discuss farm data’s place on the private-public good spectrum. For a good to be a ``private good'', i.e., its benefits and costs are fully realized by the owner, it must be both rivalrous and excludable. A ``public good'' is neither rivalrous nor excludable meaning one’s enjoyment of the good does not diminish another’s nor can anyone be prevented from using it. Farm data is unlike other farm assets in its intangibility. Copies of farm data can be shared without inhibiting its use by the original owner. In this way, farm data is clearly non-rivalrous~\cite{griffin2016property}. 

The ability of a farm operator to exclude others from using their data depends on their relationships with data service providers and the data sharing agreements that govern those relationships. For example, equipment manufacturers collect telematics data on newly sold products for the purpose of improving performance and service. The equipment owner has no reasonable expectation of excludability and may not even be aware they opted into such an agreement. Farms that subscribe to a data service provider to manage and analyze their data, e.g. Climate FieldView, are similarly forfeiting excludability. However, data may be partially excludable if access is limited to within the network, or “club,” of subscribers. As such, farm data most closely satisfies the definition of a ``club good'' \cite{miller2018estimating}.

Coble \textit{et al}. (2016)\cite{coble2016advancing} point out that a farmer’s data is not legally protected from disclosure in the way medical records are protected by the Health Insurance Portability and Accountability Act (HIPAA) or education information is protected by the Family Educational Rights and Privacy Act (FERPA). Without overarching legal safeguards for farm data, individual sharing agreements dictate the terms of access and use. Though farm data enjoy some of the intellectual property protections afforded to trade secrets, its legal ownership structure remains ambiguous~\cite{ferrell2016legal}. A recent survey by the American Farm Bureau Federation highlights this concern. Nearly 80\% of farmers reported being concerned or extremely concerned about which entities can access their data. Even among data service subscribers, 55\% did not know whether they themselves owned or controlled their data~\cite{farmbureau2016survey}. It is not surprising then that organizations such as Ag Data Transparent and the Ag Data Coalition have emerged to strengthen privacy protections for farm data and give farmers control over how their data is used. Academic research should focus on designing incentive compatible mechanisms that encourage data sharing while protecting farmers’ intellectual property.
\section{Looking Ahead}
\noindent Big data and precision agriculture will likely be a disruptive force in the farm economy over the long-term. Technological advancements have led to farm consolidation in the past. Newer and more efficient tools reduce per-unit costs, allowing small and mid-size farms to expand. Moreover, economies of scale---the ability of large firms to spread investment costs over many units of production---allow large operations to exploit precision agriculture and big data tools earlier, potentially accelerating the consolidation trend~\cite{coble2016advancing}. As mechanization continues and farm operations become more digitized, their labor needs will narrow and become specialized. This could mitigate the so-called ``brain drain'' impacting many rural communities in the U.S.~\cite{artz2003brain}. 

The ability to track virtually every farm activity means that efforts to increase sustainability can be verified and certified by third-parties. For example, Indigo Ag, a digital agriculture startup, is attempting to monetize on-farm carbon sequestration for farms using regenerative practices. Blockchain technology in particular promises to increase sustainability and ensure food safety by authenticating information across the food system. In additional to privacy concerns, adoption of precision agriculture and big data is limited by the availability of broadband internet in rural areas~\cite{whitacre2014connected}. Integrated farm equipment and sensors pipe data up to the cloud in real time meaning farmers are more reliant on upload speeds than the typical user. The societal returns to rural broadband far outweigh the private benefits accruing to rural electric cooperatives, suggesting high-speed internet will be under-supplied in the market and implying a role for external support~\cite{grant2018broadband}.

%\end{comment}

%\bibliographystyle{ieeetr}
\bibliography{ref,schaterji}
\newpage

\end{document}